\documentclass[%
    aapm,
    mph,%
    amsmath,amssymb,
    reprint,%
    superscriptaddress 
]{revtex4-2}

    \usepackage{graphicx}
    \usepackage{dcolumn}
    \usepackage{bm}
    \usepackage[mathlines]{lineno}
    \usepackage[footnote=true]{snotez}
    \usepackage{verbatim} 
    \usepackage{amsmath}
    \usepackage{soul}

\begin{document}

\preprint{AAPM/123-QED}

\title[]{In situ gas cell for the analysis of adsorption behavior on surfaces using x-ray spectroscopy }

\author{Cornelia Streeck*}
    \affiliation{Pysikalisch-Technische Bundesanstalt (PTB), Abbestr. 2-12, 10587 Berlin, Germany}

\author{Daniel Grötzsch*,**}%
    \affiliation{Technische Universität Berlin (TU-Berlin), Hardenbergstr. 36, 10623 Berlin, Germany}

\author{Jan Weser}
    \affiliation{Pysikalisch-Technische Bundesanstalt (PTB), Abbestr. 2-12, 10587 Berlin, Germany}

\author{Andreas Nutsch}
    \affiliation{Pysikalisch-Technische Bundesanstalt (PTB), Abbestr. 2-12, 10587 Berlin, Germany, current address: ASML Berlin GmbH, Waldkraiburgerstrasse 5, 12347 Berlin, Germany}
\author{Wolfgang Malzer}
    \affiliation{Technische Universität Berlin (TU-Berlin), Hardenbergstr. 36, 10623 Berlin, Germany}
\author{Burkhard Beckhoff}
    \affiliation{Pysikalisch-Technische Bundesanstalt (PTB), Abbestr. 2-12, 10587 Berlin, Germany}
\author{Birgit Kanngießer}
    \affiliation{Technische Universität Berlin (TU-Berlin), Hardenbergstr. 36, 10623 Berlin, Germany}
\author{Ioanna Mantouvalou}
    \affiliation{Helmholtz Zentrum Berlin (HZB), Albert-Einstein-Str. 15, 12489 Berlin, Germany}

\date{\today}
\begin{abstract}
    * contributed equally\\ 
    \\
    A gas cell for in-situ measurements of Volatile Organic Compounds (VOCs) and their adsorption behavior on different surfaces by means of X-Ray Fluorescence (XRF) and X-ray Absorption Fine-Structure (XAFS) spectroscopy has been developed. The cell is especially designed to allow for the efficient excitation and detection of low-Z elements such as carbon, oxygen or nitrogen as main components of VOCs. Two measurement modes are available. In the surface mode, adsorption on a surface can be studied using XAFS by fluorescence detection under shallow angles of incidence. The transmission mode enables the simultaneous investigation of gaseous samples via XAFS in transmittance and fluorescence detection modes. Proof-of-principle experiments were performed at the PTB plane grating monochromator beamline for soft x-ray radiation at the synchrotron radiation facility BESSY II. The flexible design and high versatility of the cell are demonstrated with the investigation of ethanol (EtOH) as one of the most abundant VOCs. The comparison of Near-Edge X-ray Absorption Fine-Structure (NEXAFS) spectra under transmission and fluorescence detection in the gas phase with measurements of adsorbed molecules on a Si-wafer surface both at the C and O\nobreakdash-K absorption edges proves the applicability of the cell for the monitoring of adsorption processes.
\end{abstract}
\maketitle

\section{\label{sec:intro}Introduction}
 Volatile organic compounds (VOCs) are defined as carbon-containing molecules that can easily evaporate in ambient air conditions up to temperatures of 300°C. VOCs are emitted into the atmosphere from both anthropogenic (e.g. fossil fuels or industrial solvents) and natural sources (e.g. plant metabolism) and play a key role in the photochemical formation of air pollutants and aerosols that affect climate and human health. For monitoring air quality and air pollution levels of these compounds, measurements take place in ambient and indoor air, up to measuring of vertical VOC profiles in the atmosphere to better model and monitor their impact on climate.\cite{koppmann_chemistry_2020,baldan_european_2013,dieu_hien_overview_2019, demichelis_compact_2018,englert_preparation_2018,rhoderick_stability_2019} VOC concentrations in the background atmosphere are typically in the low range of a few pmol$\cdot$mol$^{-1}$ to a few nmol$\cdot$mol$^{-1}$ and require measurement techniques with very high sensitivity that are adapted depending on the concentration and compound to be detected (Table 1 in \cite{rhoderick_stability_2019}). Gas chromatographs with flame ionisation detectors and mass spectroscopy detectors are currently the most commonly used VOC measuring instruments, but also Fourier transform infrared spectroscopy, cavity ring-down spectroscopy, portable monitors and direct sensor techniques specified for individual VOCs, e.g. for breath analysis, are used. \cite{koppmann_chemistry_2020,rhoderick_stability_2019,englert_preparation_2018,demichelis_compact_2018,baldan_european_2013} Good calibration standards are essential for metrological traceability and comparability. Direct calibration of VOC detectors in routine analyses of gas samples is performed with a VOC gas reference standard mixture prepared by gaseous VOC standard cylinders or by dynamic in situ production from certified liquid VOC standard solutions. Different containers can also be used for sampling for subsequent laboratory analysis. Besides temporal stability (for monitoring or long-term), a major challenge is the loss of VOCs due to adsorption on surfaces, which leads to poor long-term stability, memory effects that are difficult to calculate and thus to a high uncertainty, especially for reactive VOCs and the concentrations occurring in the trace range. In order to minimise the interaction with the surfaces, the state of research also includes the use of different surface treatments of the containers, gas cylinders and the components used, such as filters and pipe systems in the measurement setup.\cite{dieu_hien_overview_2019,baldan_european_2013, demichelis_compact_2018,englert_preparation_2018, li_advanced_2019,rhoderick_stability_2019} The development of the gas cell presented here is linked to this issue.
 \\
While it is interesting to understand the adsorption behavior fundamentally and quantify adsorbed material, also the adsorption mechanisms can be utilized as a means of reducing the amount of pollutants in the atmosphere. For this purpose, not only the verification of adsorption of VOCs on different surfaces is mandatory but also the analysis of the dynamics involved. To gain access to the number of adsorbed molecules on a surface, a gas cell was developed for the investigation of the main constituents of VOCs such as carbon (C) or oxygen (O) in the spectral range of soft x\nobreakdash-ray radiation. X\nobreakdash-ray fluorescence spectroscopy is a suitable non-destructive analytical tool which enables quantitative element-specific investigations and when combined with x\nobreakdash-ray absorption spectroscopy facilitates the elucidation of the chemical environment of an analyte, e.g. the molecular orbitals, oxidation states or binding angles.
\\
X\nobreakdash-Ray Fluorescence (XRF) occurs as one of the possible relaxation processes after an ionization of an inner-shell electron by incident x\nobreakdash-ray radiation. Characteristic fluorescence photons of the excited element can be detected and give insights about the chemical composition of a sample. The characteristic x-ray radiation can be used for quantitative analysis of elemental fractions based on the evaluation of the detected intensity.\cite{tertian_principles_1982} A very surface sensitive adaptation of the XRF technique with detection limits in the ppm or even ppb range is the total reflection x\nobreakdash-ray fluorescence (TXRF) analysis.\cite{von_bohlen_total_2009,unterumsberger_interaction_2020,beckhoff_reference-free_2007, unterumsberger_round_2021} Even for organic molecules this technique can be applied to determine the surface functional group densities, e.g. shown for silanized surfaces.\cite{dietrich_quantification_2015,fischer_quantification_2015} 
Here, the characteristic x\nobreakdash-ray radiation is excited under very shallow angles (below the critical angle of x\nobreakdash-ray reflection) with an angle of detection of about 90° facilitating a minimal distance between energy-dispersive detector and sample surface. If the surface is sufficiently smooth, the excitation beam can interfere with the reflected beam forming an X\nobreakdash-ray Standing Wave field (XSW), which modulates the excitation efficiency. This XSW field can excite elements deposited at the surface with up to twice the intensity of the incident beam while nearly no background signals from the substrate disturb the x\nobreakdash-ray fluorescence line of interest. A small sample-detector distance results in a large solid angle of detection allowing for high counting statistics of the recorded fluorescence radiation. This TXRF methodology is implemented in the developed in-situ gas-flow cell. An exchangeable substrate in a gaseous atmosphere can be irradiated in TXRF-geometry and the interaction between substrate surface and gas is analyzed during gas flow as a function of time.  
\\
The second methodology addressed with the novel gas-flow cell is the X-ray Absorption Fine Structure (XAFS) spectroscopy. By tuning the excitation photon energy over an absorption edge of an element of interest with high energy resolution detected, and observing either the overall transmission or the fluorescence intensity of a specific line the fine structure in the edge region can be investigated. The analysis of this fine structure gives information about (molecular) bonds and bond distances. The gas-flow cell facilitates XAFS experiments in transmittance using a photodiode and XAFS in fluorescence detection mode by using an energy-dispersive SDD detector for an characteristic fluorescence lines of an element of interest.
\\
The ultra-high to high absorption of soft x-rays in air requires the use of vacuum equipment in the entire measurement setup, from the beam tube of the measurement chamber to the detector. This means that the gas cell, which enables a sample environment for the beam in this vacuum setup, must not only have short path between entrance and exit and the possibility to regulate the gas flow, but also windows with a high transmittance that separate the gas from the rest of the setup. In the novel gas cell, thin silicon nitride (SiN) windows with a thickness of 150~nm are carefully mounted which show a transmission of 0.2 to 0.6 in the range between 250~eV and 600~eV which contains the K absorption edges of C, N and O. 
\\
There are several studies in literature dealing with XAFS measurements on gaseous samples. One straight forward approach is presented by Nakanishi et al.\cite{nakanishi_xafs_2010}, where the gaseous sample is introduced directly into the measurement chamber which is seperated from the incoming beam by means of a Be~window. An et al.\cite{an_facile_2014} demonstrate the applicability of a straightforward heating cell for in-situ transmittance and fluorescence XAFS measurements and Yao et al.\cite{yao_unraveling_2014} introduce a cell for simultaneous use with x-ray and IR irradiation. Both these cells are suited for experiments with x-ray energies above 2 keV. 
\\
Guo et al.\cite{guo_development_2013} describe a cell with a similar excitation and detection geometry as the cell presented in this work without the possibility to investigate surfaces. A cell for the investigation of catalytic processed by Aguilar-Taipa et al.\cite{aguilar-tapia_new_2018} facilitates transmission and fluorescence measurements only in the hard x-ray regime due to the use of Be windows. In this cell a reaction area is implemented in the center for in-situ experiments. Benkert et al.\cite{benkert_setup_2014} present a gas cell designed to study the electronic structure of gases and gas/solid interfaces using x-ray emission and absorption techniques. This is implemented by using one window for excitation and detection with the advantage of a very small distance between window and sample surface ($<$ 100 µm). A reaction cell for ambient pressure soft x-ray absorption spectroscopy by Castan-Guerrero et al.\cite{castan-guerrero_reaction_2018} utilizes thin SiN windows, in which the detection is performed by means of total electron yield measurement. 
\\
In this work, the cells design is presented and the experimental modi described. As a proof-of-principle experiment, the adsorption of ethanol (EtOH) as one of the most abundant VOCs on a silicon wafer surface is investigated using x-ray absorption spectroscopy, showing the feasibility of detection and chemical speciation of minute amounts of VOCs after adsorption.  
\section{\label{sec:design}Design}
    \begin{figure}
    \includegraphics[width=0.47\textwidth]{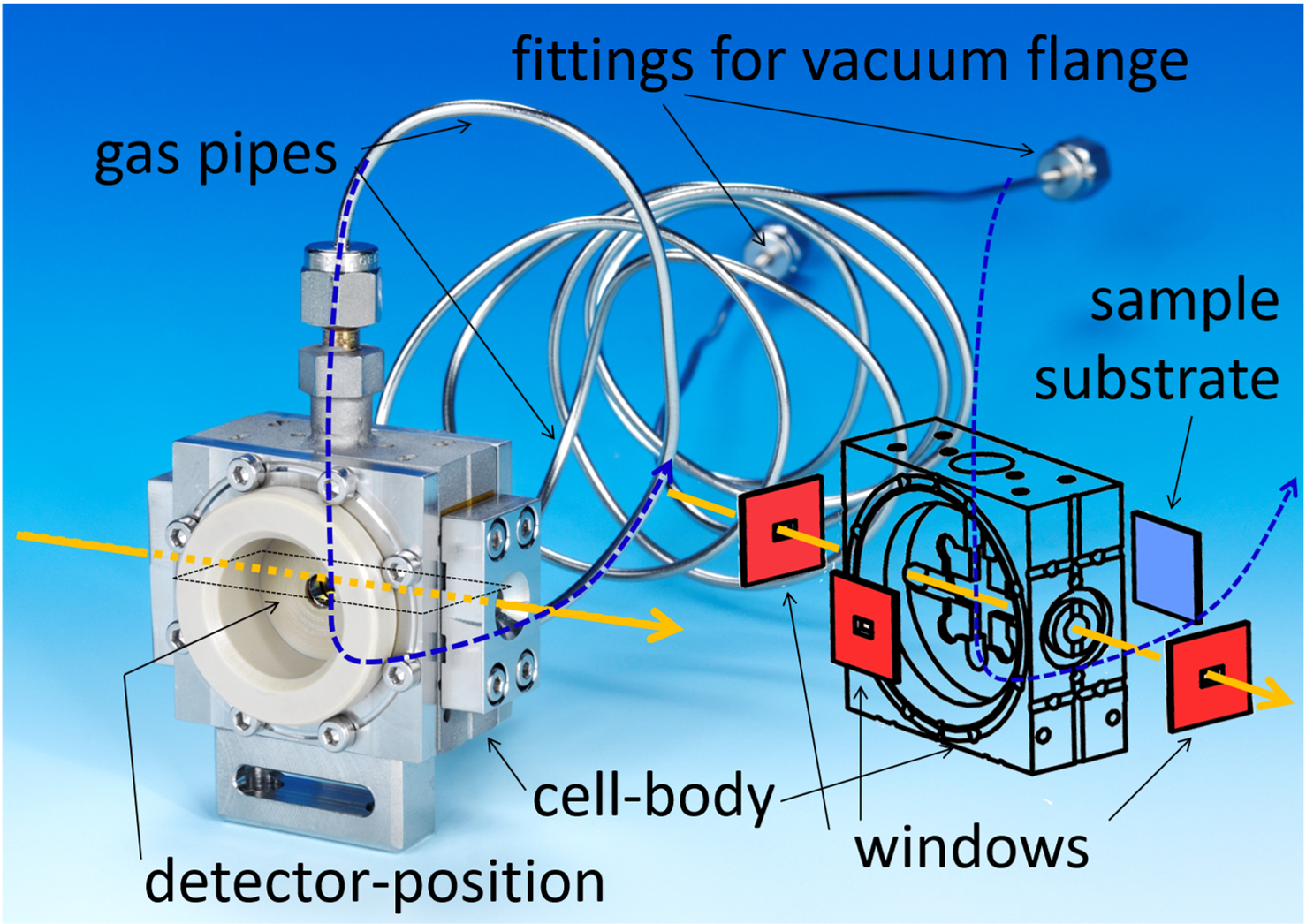}
    \caption{\label{fig:gascell}Photo of the mounted gas-flow cell (left and technical drawing of the body and the three exchangeable SiN windows and samples (right). The yellow beam depicts the beam path for a transmission experiment through two of the windows allowing the measurement of the transmitted beam intensity. Through the third window the characteristic x-ray fluorescence radiation can be collected either from the gas or the substrate surface by an x-ray detector.}
    \end{figure}
The gas-flow cell was developed to allow for the analysis of low-Z constituents of VOCs in a gas-ambience and their adsorption behavior with respect to different surfaces. Due to the intended use of soft x-rays, operation in an ultra-high vacuum (UHV) chamber is mandatory. The design of the gas-flow cell accomplishes on the one hand side an enclosed volume for a gas-flow with an adjustable gas pressure between $0.1$~mbar and ambient atmosphere separated from the vacuum conditions in the UHV chamber, and on the other hand allows for soft x-ray experiments by using ultra-thin windows with sufficient transmittance. Due to the modular design and compactness of the novel gas cell presented, it can also be transferred to other beamlines and x-ray sources to be installed in existing vacuum chambers.\\
    \begin{figure*}
    \includegraphics[width=0.9\textwidth]{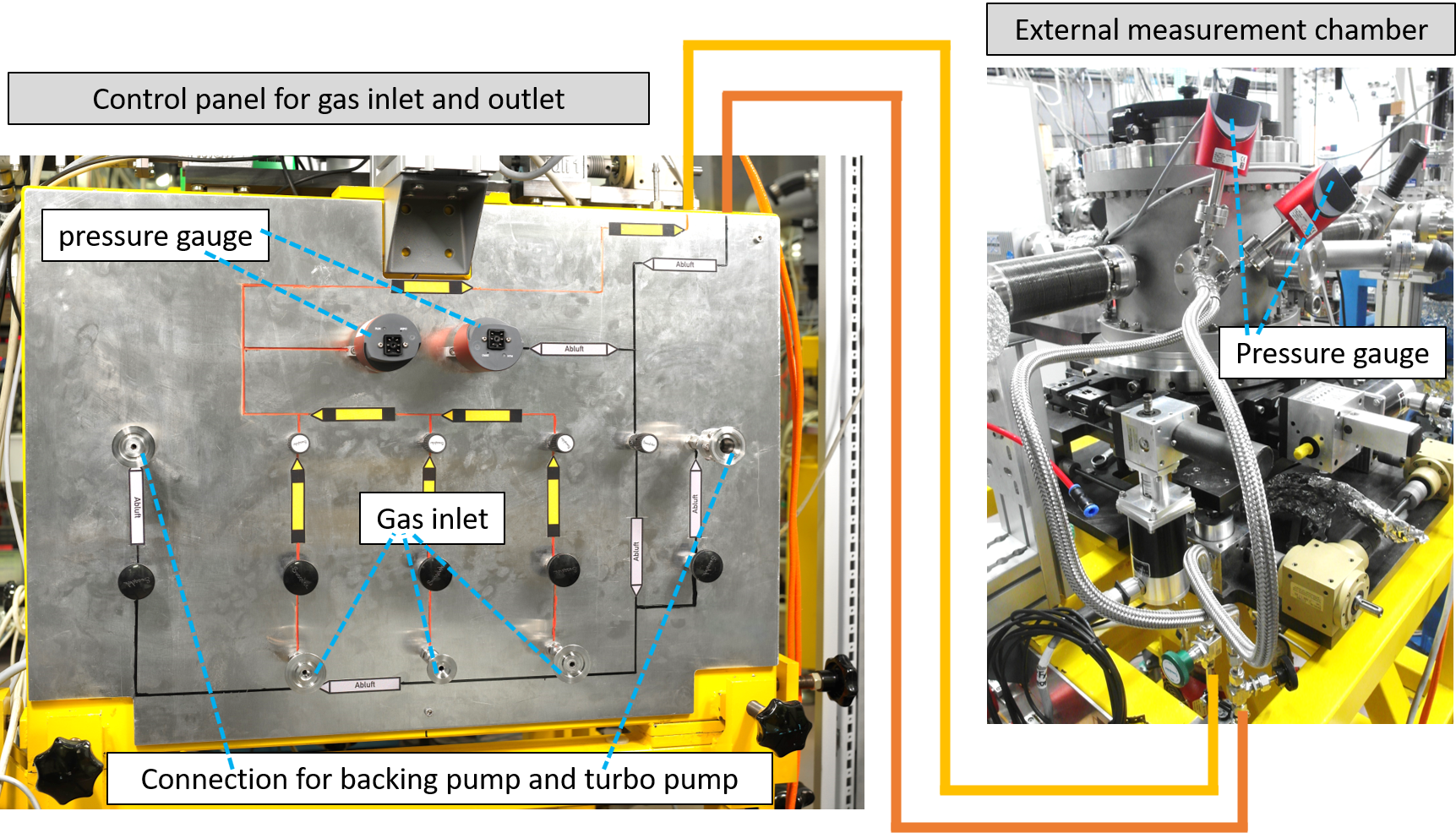}
    \caption{\label{fig:control-panel}The gas cell is operated in the experimental chamber (right) and the gas pipes are conducted through a  vacuum flange. For a versatile pumping and gas inlet a control panel allows for the parallel installation of three different gases and two pump stations. Directly at the vacuum flange and near the gas inlet/outlet pressure gauges are installed for monitoring.}
    \end{figure*}
The cell has a compact stainless-steel housing, which is vacuum-tight with a gas-inlet and outlet and three exchangeable ultra-thin windows which transmit the soft x-ray radiation, see Fig.~\ref{fig:gascell}. The gas flow through the gas cell is realized using inlet and outlet gas pipes (see Fig.~\ref{fig:control-panel}) which can be connected through the wall of the experimental chamber by a swage lock connector attached to a vacuum flange. In this way, the gas cell can be pumped and flooded independently from the vacuum chamber. The pipes outside of the experimental chamber are connecting a backing pump and a turbo pump at the outlet and three parallel inlets separated by valves. Here, a nitrogen bottle (for ventilation and flushing) and the container with the VOC can be attached at the same time. In addition, the inlet side can also be pumped separately. The pressure is monitored by two pairs of capacitance pressure gauges from Pfeiffer Vacuum, which provides measurements that are independent of the type of gas (see Fig.~\ref{fig:control-panel}). One pair is directly mounted behind the vacuum flanges of the gas cell pipes at the measurement chamber (CRM 362, measuring range from $10^{-2}$ mbar - $110$ mbar) and one pair is mounted near the gas inlet/outlet (CRM 361, measuring range from $10^{-1}$ mbar - $1100$ mbar). These pressure sensors on the gas pipes provide an indirect information about the pressure in the cell. Window breakage can be prevented also by monitoring the pressure. This is especially important when the pressure difference between the inside of the cell and the measuring chamber is large during venting and pumping.
\\
The cell allows for two different measurement geometries – measurements in 'surface mode' and in 'transmission mode' of operation. In the surface mode, adsorption on a surface can be studied using XAFS by fluorescence detection under shallow angles of incidence. The transmission mode enables the simultaneous investigation of gaseous samples via XAFS in transmittance and fluorescence detection modes. At three positions of the gas cell wafer-based SiN windows with a frame size of (10 x 10)~mm² (depicted in red in Fig.~\ref{fig:gascell} and W1-W3 in Fig.~\ref{fig:meas-modes}) can be mounted.  The sample substrate (blue plate in  Fig.~\ref{fig:gascell} and shown in Fig.~\ref{fig:meas-modes}) for the VOC-adsorption experiment can be exchanged with maximal dimensions of (10 x 10 x 1)~mm³. 
\\
For x-ray fluorescence detection an energy-dispersive windowless Silicon-Drift-Detector (SDD) with a chip size of about 16~mm² is used. The calibrated SDD detector has in the energy range of the C, N and O\nobreakdash-K fluorescence lines a detection efficiency between 0.6 to 0.9 in this energy range. With the help of a guidance bar, the detector can be moved up to 3~mm through the cell housing (see Fig.~\ref{fig:gascell}) directly in front of the third window for the fluorescence detection (W3 in Fig.~\ref{fig:meas-modes}). This is realised by a vacuum bellows to wind the detector forwards and backwards to the window W3 of the gas cell. The gas cell can be modular equipped with different cap attachments to guide detectors with different diameters. This minimal distance between cell and detector optimizes the solid angle of the detection. For the transmission measurements a calibrated photodiode behind the gas-cell is available.
    \begin{figure}
    \includegraphics[width=0.47\textwidth]{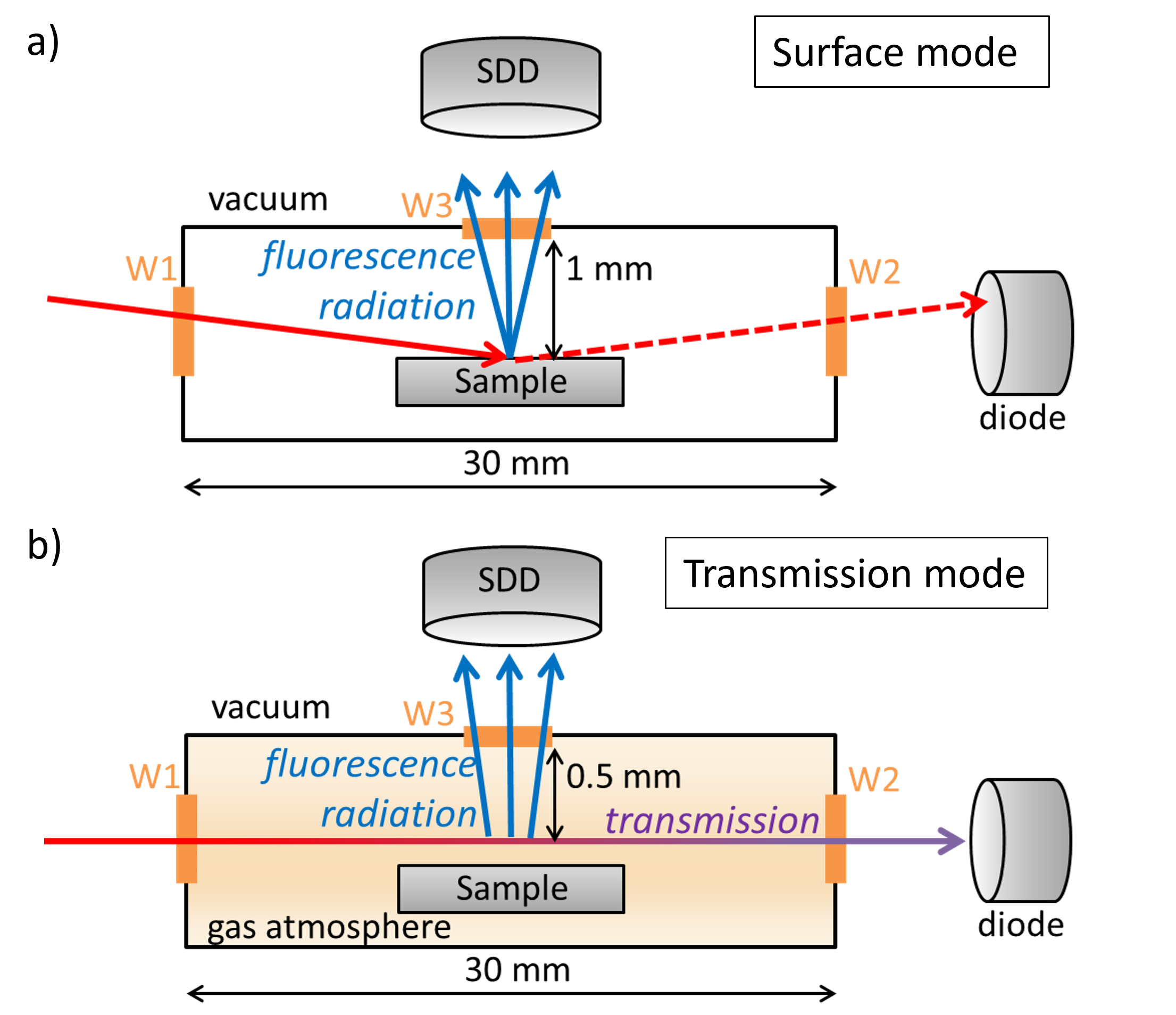}
    \caption{\label{fig:meas-modes}Both possible excitation geometries for the operation of the gas-flow cell are depicted. a) surface mode of operation, b) transmission mode of operation. For further descriptions see the text.}
    \end{figure}
\subsection{\label{sec:Surf}Surface mode of operation}

In Fig.~\ref{fig:meas-modes} (a) the surface mode of operation when using the gas-flow cell is depicted: The sample surface is excited under a shallow angle leading to an enlarged footprint on the sample surface. The characteristic fluorescence radiation emitted from the sample surface is detected through window W3. Depending on the angle of incidence it is possible to use the photodiode behind the chamber to monitor the x-ray reflection signal from the surface.
\\
For a quantitative analysis of the mass deposition of element $i$ on the surface the photon flux $I_{i,j}^*$ of the characteristic fluorescence line $j$ of element $i$ can be calculated using the following equation:\cite{sherman_theoretical_1955,mantler_quantitative_2006,unterumsberger_complementary_2011, unterumsberger_interaction_2020}
        \begin{eqnarray}
        \label{eq:surface}
        I_{i,j}^{*,surf}=\frac{I_0^*Q(E_0)}{\text{sin}\Psi_{in}} \frac{\Omega_{det}}{4\pi}\frac{\tau_{X_i}}{\mu_{tot,i}} \left(%
        1-e^{-\mu_{tot,i}\frac{m_i}{F}}
        \right)%
        ,
        \end{eqnarray}
with the total mass attenuation coefficient
$$\mu_{tot,i}=\frac{\mu_{i,E_0}}{\text{sin}\Psi_{in}}+\frac{\mu_{i,E_{i,j}}}{\text{sin}\Psi_{out}}$$ 
and the fluorescence production cross section $$Q(E_0)=\omega_{X_i}g_{j,X_i}.$$
Here, $\mu_{i,E_0}$ and $\mu_{i,E_{i,j}}$ are the mass attenuation coefficients for element $i$ at the excitation energy $E_0$ and the fluorescence line energy $E_{i,j}$, which are related to the sine of the incident angle $\Psi_{in}$ and the angle of observation $\Psi_{out}=90$°$-\Psi{in}$. For mixtures of several elements instead of a single-element sample $i$, $\mu_{tot,i}$ must be calculated according to the mass fraction of each element.\cite{sherman_theoretical_1955,mantler_quantitative_2006} Furthermore, in Eq. (\ref{eq:surface}) $\omega_{X_i}$ is the fluorescence yield of the absorption edge $X_i$ and $g_{j,X_i}$ is the transition probability of the fluorescence line $j$. $\tau_{X_i}$ is the partial photoelectric cross-section of the absorption edge $X_i$ and as the mass attenuation coefficient it depends on the excitation energy $\tau_{X_i}=\tau_{X_i} (E_0)$.\\
The photon flux of the incident radiation $I_0^*$ and the emitted photon flux $I_{i,j}^*$ are marked with an asterisk to emphasize that the window transmissions for the incident beam as well as for the detected count rate through the windows W1 and W3 must be taken into account. In addition, for very shallow incident angles the incoming photon flux $I_0^*$ has to be corrected by the relative intensity of the XSW field which occur on smooth surfaces.\cite{von_bohlen_total_2009} Furthermore, in Eq. (\ref{eq:surface}) $\Omega_{det}$ is the solid angle of detection and $m_i/F$ the mass deposition of element $i$ with the unit area $F$. With all relevant instrumental and experimental parameters as well as the the respective fundamental parameters\cite{schoonjans_xraylib_2011} known, a reference-free quantification of $m_i/F$ can be implemented.\cite{unterumsberger_interaction_2020,beckhoff_reference-free_2007,unterumsberger_round_2021} Then, the window transmissions and the geometric conditions (projected beam profile and angular acceptance in the detection channel) for the solid angle of detection must be carefully determined and taken into account for the gas cell.
\subsection{\label{sec:trans}Transmission mode of operation}

In Fig.~\ref{fig:meas-modes} (b) the transmission mode of operation for the gas-flow cell is depicted: In this mode the gas-cell is aligned with respect to the incoming beam so that the sample surface is not irradiated and the transmission through both windows W1 and W2 and the gaseous volume can be directly measured by the photodiode behind the gas-flow cell. In addition to the transmitted radiation, the characteristic fluorescence radiation from the excited gas can be detected through window W3 with the energy-dispersive SDD. This application mode can be particularly interesting for the gas pressure determination or alternatively, the determination of the fluorescence yield of the gas atoms.
To ensure transmission in the soft x-ray region, especially at the O\nobreakdash-K edge or C\nobreakdash-K edge, the pressure in the gas cell has to be in the mbar range. At these pressures and photon energies, secondary excitation processes can be excluded. For the description of the detected signal at the photodiode behind the cell, Lambert Beers law can be applied directly, linking the transmitted intensity $I_{trans}$ to the attenuation $\mu \varrho d$ with $\mu$ as mass attenuation coefficient, $\varrho$ as mass density of the transmitted gas and $d=30$~mm the length of the gas-cell. 

        \begin{eqnarray}
        \label{eq:lambert-beer}
        I_{trans}=I_0^{\star}  e^{-\mu_{gas}\varrho d}
        \end{eqnarray}
Note, $I_0^{\star}$ differs from $I_0^*$ in Eq. (\ref{eq:surface}) by an additional window transmission through window W2. Assuming an ideal gas the density $\varrho$ can be correlated to the pressure $p$ of the gas volume by the ideal gas law  
$$\varrho=p \frac{M}{RT}$$
with the molar mass $M$, the ideal gas constant $R$ and the temperature $T$.\\
At the SDD, the detected fluorescence intensity $I_{i,j}^{*,gas}$ of the excited gas volume through W3 varies slightly with respect to Eq.~(\ref{eq:surface}) as it has to be adapted for the special geometry:
        \begin{eqnarray}
        \label{eq:gas-trans}
        I_{i,j}^{*,gas}={I_0^{**}Q(E_0)}\frac{\Omega_{det}}{4\pi}\frac{W_i \tau_{X_i}}{\mu_{gas}} \left(%
        1-e^{-\mu_{gas}\varrho y}
        \right)%
        ,
        \end{eqnarray}
Some simplifying assumptions were made: $I_0^{**}$ is constant over $y$, the length of the field of view of the SDD. The finite size of the beam perpendicular to the direction of beam propagation was neglected in the formula. In contrast to Eq. (\ref{eq:surface}), the double asterisk for $I_0^{**}$ in Eq. (\ref{eq:gas-trans}) indicates that the weakening of $I_0^*$ for half the length $d$ within the gas-cell has to be taken into account.

\section{\label{sec:experimental}Experimental}
The experiments were carried out at the plane grating monochromator beamline\cite{senf_plane-grating_1998} for undulator radiation in the laboratory of the Physikalisch-Technische Bundesanstalt\cite{beckhoff_quarter-century_2009} at the electron storage ring BESSY II in Berlin, Germany. The beamline provides soft x-ray radiation in the photon energy range from 78~eV to 1860~eV. The NEXAFS measurements for the C\nobreakdash-K edge were performed between 250~eV and 350~eV and for the O\nobreakdash-K edge between 490~eV and 590~eV, respectively. To optimize measurement time, the energy step size was varied between 0.2~eV and 5~eV at the C\nobreakdash-K edge and between 0.25~eV and 20~eV at the O\nobreakdash-K edge, resulting in an overall measurement time of 24 min and 14 min, respectively by using a life time preset of 10~s and 7~s. As the Peltier-cooled SDD is windowless for an optimized efficiency in the soft x-ray region, a pressure of $10^{-6}$~mbar or better must be ensured in the vacuum chamber, where the detector is attached. For comparison, the minimal pressure in the cell was approx. $0.1$~mbar and the maximum pressures were below 50~mbar.
The gas cell was installed at the focus point of the pgm beamline. At the focus, the beam size amounts to approximately 150~µm full width half maximum (FWHM) in the horizontal plane. With the exit slit of the monochromator a beam size of 40~µm FWHM in the vertical plane was used. With the selected beamline settings, an energy resolving power of $E/\Delta E=2000$ at the C\nobreakdash-K edge can be achieved.\cite{mantouvalou_single_2016}     
\\
For the VOC adsorption experiments, SiN windows with an area of (1 x 2)~mm² and a thickness of about 150~nm were used. For transmission measurements the gas cell is adjusted with a photodiode so that the incident beam passes through the center of the gap between the sample surface and the SiN window W3. This ensures that for flat specimens which are inserted in the sample position, the surface is parallel to the incidence beam. To measure in surface mode geometry, the whole experimental set up (measurement chamber with included cell in the center) is rotated, in the presented experiments by 1°. Due to slight tolerances in the manufacturing of the window holders, an uncertainty in the angular positioning of 0.3° is assumed.  

\section{\label{sec:results}Results}

\subsection{\label{sec:trans-result}Transmission mode: comparison of XAFS recorded in transmittance and fluorescence detection}

\begin{figure}
    \includegraphics[width=0.47\textwidth]{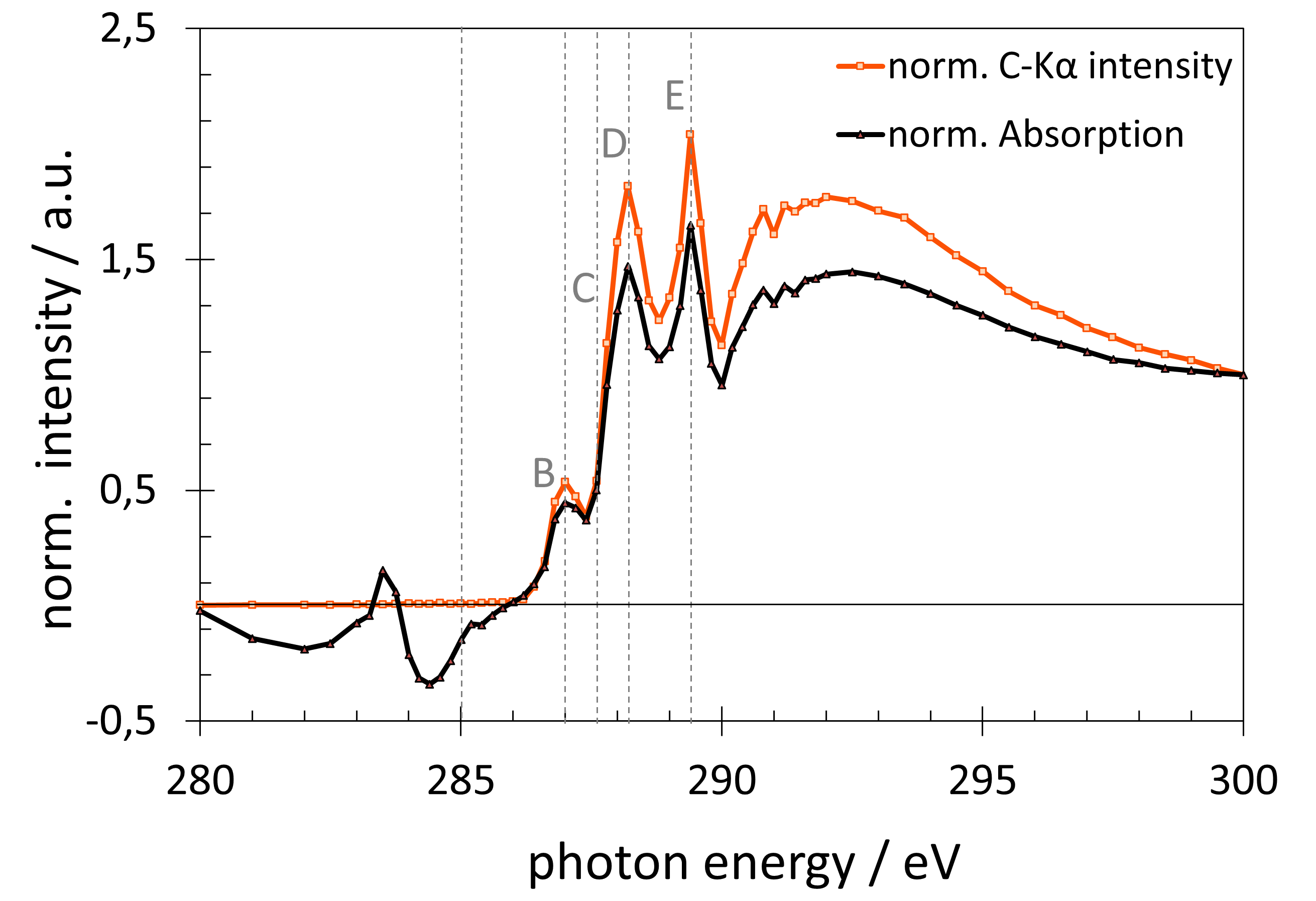}
    \caption{\label{fig:result_gasl}C-K edge spectra of gaseous EtOH: shown are the spectra obtained simultaneously from a measurement in transmission mode using the photodiode for the partial absoption spectrum of the C-K shell (black) and the SDD for the C-K$\alpha$ fluorescence line intensity (orange). The energy positions B-E mark line positions for later fitting and comparison with surface mode measurements.}
    \end{figure}

The first proof-of-principle experiment was performed in transmission mode with gaseous EtOH in order to compare the agreement XAFS in transmittance and fluorescence detection at the C\nobreakdash-K absorption edge. This demonstrates the merits of the simultaneous collection of both spectra.   
Here, due to the different detection modes the agreement between the shape of the spectra highly depends on normalization procedures. For the transmission measurements, the transmission characteristics of the two SiN windows must be taken into account. These can be obtained through a calibration measurement with a completely evacuated cell. In the case of the fluorescence detection through window 3, which occurs in approx. 90° to the incoming beam, the excitation intensity $I_0$ from the transmission measurements can only be used for half the distance (window W1 to middle of the cell). A further factor which has to be considered is, that minute adsorption of C-containing molecules on the SiN windows may occur during the measurement.
Additionally, to compare fluorescence detection and transmission XAFS data directly, the absorption of the lower shells must be subtracted from the transmission curve. For more information on this substraction data procedure see Unterumsberger et al.\cite{unterumsberger_accurate_2018} Equ. (5) and Fig. 2, and also Fig. 2 in Hönicke et al.\cite{honicke_experimental_2016}.
\\
After evacuation of the gas cell, it was filled with gas from a reservoir of evacuated liquid EtOH. A working pressure of 5~mbar was adjusted with 1 percent stability during the measurement. Fig.~\ref{fig:result_gasl} shows the C\nobreakdash-K edge absorption spectra collected simultaneously in transmission and fluorescence normalized to the value at 300~eV for comparison. Both spectra show the same distinct features of the gaseous EtOH and follow the same general trend.
The energy positions B-E plotted in Fig.~\ref{fig:result_gasl} mark distinct features and line positions  which all spectra were fitted for comparison with the surface mode measurements to prove EtOH adsorption(see next subsection and Table~\ref{tab:C-edge}).
Due to the minute number of molecules in the cell and the normalization on the window-transmission, the transmission spectrum of EtOH in black shows artefacts below 285~eV. These artefacts are not visible in the fluorescence signal (orange) due to the restriction to the C\nobreakdash-K$\alpha$ fluorescence line for NEXAFS evaluation. Only the C\nobreakdash-K$\alpha$ fluorescence photons created in the field of view of the SDD detector above the K\nobreakdash-edge with an excitation energy of 286~eV contribute to the NEXAFS signal in fluorescence detection. The artefacts in transmission mode are caused by C depositions on the windows during measurements and  modulate the overall absorption behaviour. This deposition in the measurement process needs to be extracted by a better calibration and normalisation routine. Furthermore, no self-absorption effects are visible in the fluorescence spectrum due to the low EtOH pressure in the cell. 
    \begin{figure*}
    \includegraphics[width=1.0\textwidth]{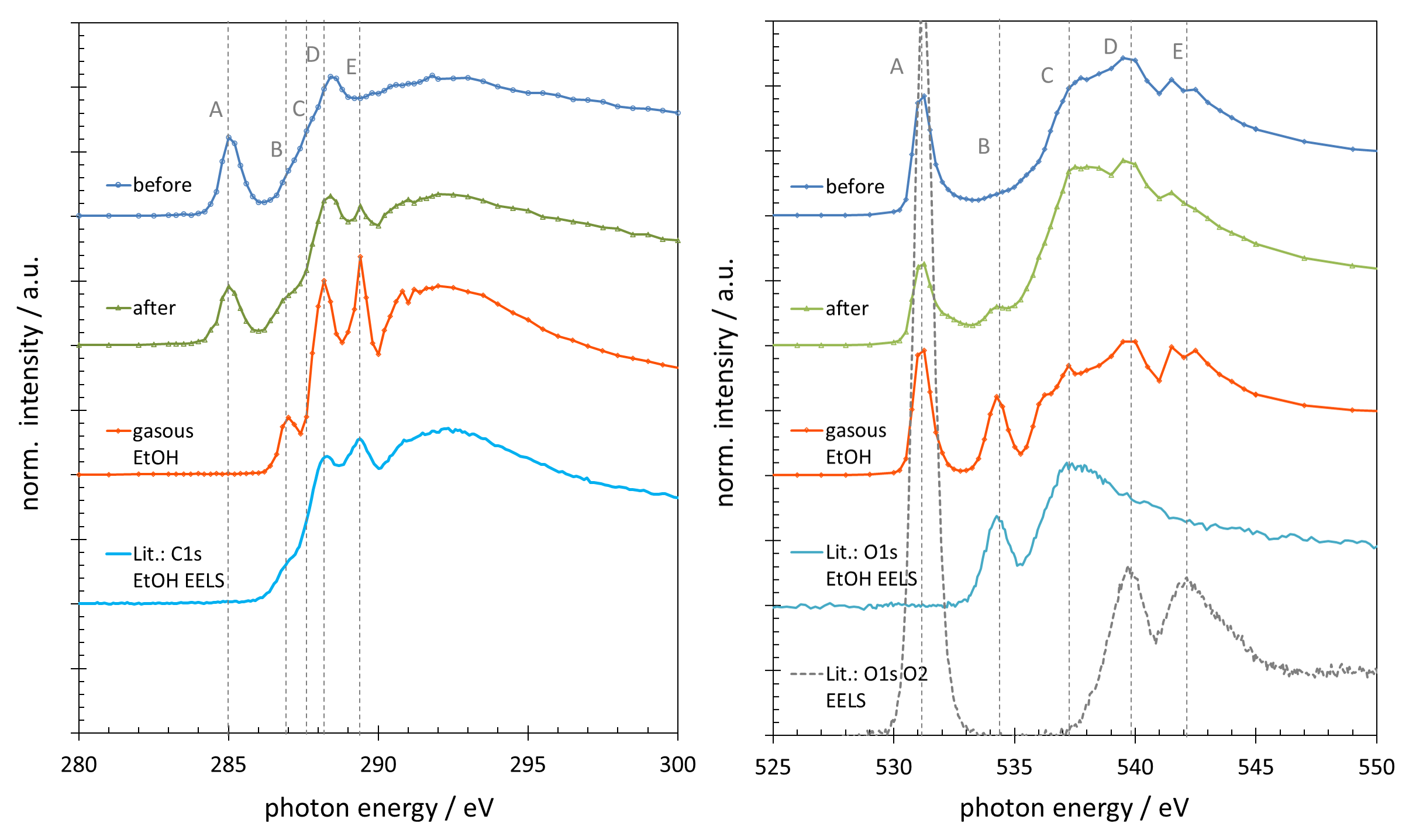}
    \caption{\label{fig:result_surfl}NEXAFS spectra at the C and O-K edge: NEXAFS spectra under fluorescence detection from the surface before (blue) and after (green) EtOH treatment are depicted.  Additionally, for comparison,  the transmission mode gas-phase measurements in orange and EELS literature reference data in light blue and grey are depicted. The energy positions of the marked features were determined in a fit routine using single and multiple Gaussian functions (see Table~\ref{tab:C-edge} and Table~\ref{tab:O-edge}). For further information see the text.}
    \end{figure*}
\subsection{\label{sec:surf-result}Surface mode: EtOH detection on surface}

As a next step the cell was operated in surface mode to test the sensitivity for adsorption by detection of EtOH molecules on a Si-wafer surface. First, XAFS under fluorescence detection were taken from the cleaned wafer-surface at the C and O\nobreakdash-K absorption edge, see the blue ‘before' curves at the top in Fig.~\ref{fig:result_surfl}. Then, the cell was flooded with 20~mbar of EtOH and evacuated to $0.3$~mbar again. The minimum pressure of 0.1~mbar could not be reached in a reasonable pumping time. The XAFS surface-mode measurements were repeated, resulting in the green 'after' curves in Fig.~\ref{fig:result_surfl}. Additionally, for comparison,  the transmission mode gas-phase measurements in orange and literature data in light blue and grey are depicted in Fig.~\ref{fig:result_surfl}. The gas phase literature spectra of EtOH and O$_2$ were obtained from the gas phase core excitation database \cite{hitchcock_bibliography_1994,gas-phase-core-excitation-data-base_httpunicornmcmastercacorexcedb-titlehtml_2022} which contains freely available data sets of spectra recorded by dipole regime Electron Energy Loss Spectroscopy (EELS). The C1s and O1s spectra of EtOH are depicted in light blue and the O1s spectrum of O$_2$ is depicted in grey in Fig.~\ref{fig:result_surfl}. The EELS  spectra shows a lower energy resolution than the presented measurements. In particular for the O1s spectrum (grey) of O$_2$ in the region of feature E, different resonances overlap, comparable to higher resolution measurements in literature.  \cite{ma_high-resolution_1991,hitchcock_k-shell_1980,hu_near_2004}
\\
When comparing the spectra in Fig.~\ref{fig:result_surfl}, it can be seen that features B and E for the C\nobreakdash-K edge and feature B for the O\nobreakdash-K edge can be used as evidence for EtOH adsorption on the silicon wafer surface. In order to elaborate this adsorption of EtOH on the silicon wafer surface, all spectra have been fitted with a mixture of Gaussian functions and an arctan function for the edge in order to assign energies to the distinct features depicted in Fig.~\ref{fig:result_surfl}.
When setting up the fit routine, a major focus was put on achieving a stable fit for the different spectra (this work and EELS literature data), although different energy resolutions and strongly differing step sizes were used. For most of the features shown in Fig.~\ref{fig:result_surfl}, the application of a single Gaussian function was sufficient. For features A and E at the O\nobreakdash-K edge, two and three Gaussian functions, respectively, were necessary and the energy position given was determined by the centre of mass. The Gaussian functions used were based on known resonances for the transitions of molecular orbitals that are relevant here.
\cite{ma_high-resolution_1991,hitchcock_k-shell_1980, hu_near_2004,  wight_k-shell_1974, crapper_nexafs_1987, stohr_near_1987, outka_nexafs_1987, outka_curve_1988, sham_k_1989, stohr_nexafs_1992} 
The results for the features depicted in Fig.~\ref{fig:result_surfl} are listed in Table~\ref{tab:C-edge} and~\ref{tab:O-edge} for both edges. 
    \begin{table} 
    \caption{\label{tab:C-edge} C-K edge}
    \begin{ruledtabular}
    \begin{tabular}{ccccc}
    
    Peak 
    & 
    \begin{tabular}[c]{@{}c@{}}surface \\ before \\ EtOH \\ treatment\end{tabular} 
    & 
    \begin{tabular}[c]{@{}c@{}}surface \\ after \\ EtOH \\ treatment\end{tabular}
    &
    \begin{tabular}[c]{@{}c@{}}gaseous \\ EtOH\end{tabular} 
    & 
    \begin{tabular}[c]{@{}c@{}}C1s \\ EELS \\ Spectrum \\ of \\ EtOH\cite{hitchcock_bibliography_1994,gas-phase-core-excitation-data-base_httpunicornmcmastercacorexcedb-titlehtml_2022}\end{tabular} \\
    
    \hline
     & \multicolumn{4}{c}{\textbf{peak position / eV}} \\
    \hline
    A & 285.1     & 285.0     & --      & --      \\
    B & 287.1     & 287.0     & 287.0   & 287.1    \\
    C & 287.8     & 288.0     & 288.0   & 288.0    \\
    D & 288.6     & 288.5     & 288.4   & 288.6    \\
    E & 289.5     & 289.5     & 289.4   & 289.4  
                                                                          
    \end{tabular}
    \end{ruledtabular}
    \end{table}

    \begin{table} 
    \caption{\label{tab:O-edge} O-K edge}
    \begin{ruledtabular}
    \begin{tabular}{cccccc}
    
    Peak 
    & 
    \begin{tabular}[c]{@{}c@{}}surface \\ before \\ EtOH \\ treatment\end{tabular} 
    & 
    \begin{tabular}[c]{@{}c@{}}surface \\ after \\ EtOH \\ treatment\end{tabular} 
    & 
    \begin{tabular}[c]{@{}c@{}}gaseous\\ EtOH\end{tabular} 
    & 
    \begin{tabular}[c]{@{}c@{}}O1s \\ EELS \\ Spectrum \\ of \\ EtOH \cite{hitchcock_bibliography_1994,gas-phase-core-excitation-data-base_httpunicornmcmastercacorexcedb-titlehtml_2022} \end{tabular} 
    & 
    \begin{tabular}[c]{@{}c@{}}O1s \\ EELS \\ Spectra \\ of  \\ O$_2$\cite{hitchcock_bibliography_1994,gas-phase-core-excitation-data-base_httpunicornmcmastercacorexcedb-titlehtml_2022}\end{tabular} \\ 
    
    \hline
     & \multicolumn{4}{c}{\textbf{peak position / eV}} & \textbf{} \\ 
     \hline
    
    A & 531.3   & 531.4     & 531.3 & --    & 531.3 \\
    B & --      & 534.1     & 534.3 & 534.3 & -- \\
    C & 537.8   & 537.6     & 537.3 & 537.5 & -- \\
    D & 539.8   & 539.8     & 539.8 & 539.8 & 539.7 \\
    E & 542.4   & 542.4     & 542.3 & 542.1 & 542.4
    
    \end{tabular}
    \end{ruledtabular}
    \end{table}

At the C\nobreakdash-K edge from the two distinct features at A and D in the dark blue 'before' spectrum in Fig.~\ref{fig:result_surfl}, a C contamination on the Si surface can be observed.\cite{mangolini_quantification_2021,sinha_study_2018} After EtOH treatment the green 'after' spectrum shows two more features B and E wich can be assigned to EtOH adsorption.
\\
At the O\nobreakdash-K edge the dark blue spectrum 'before' shows the native SiO$_2$ layer of the substrate which is superimposed with the spectrum of O$_2$ from residual air in the cell\cite{ma_high-resolution_1991, hu_near_2004,hitchcock_k-shell_1980}, which can be deduced from the previous calibration measurement of the 'evacuated' gas cell. Especially the features A, D and E are dominated by the residual oxygen.
The ‘after’ measurement in green at the O\nobreakdash-K edge is then composed of three characteristic spectra, the SiO$_2$ spectrum, an EtOH contribution (feature B) and also a pure O$_2$ contribution stemming again from residual air in the cell. \\
For both experiments, when evaluating the surface-mode measurements, the extent to which the EtOH content in the residual gas has an influence on the signal level must be taken into account for the evaluation of EtOH on the silicon wafer surface, as a minimum pressure of approx. 0.3~mbar could be achieved for the measurements of the surface contamination.  
As an estimation of the effect, the absolute count rates of the C\nobreakdash-Ka and O\nobreakdash-Ka events of the SDD spectra, normalized to the incident photon number $I_0^{**}$, were compared: on the one hand, the signal height from the transmission gas measurement and, on the other hand, from the surface measurements. In a first approximation, we assume that the excited gas volume of both measurements is comparable and that the pressure difference between both measurements (Transmission mode approx. 5~mbar, Surface mode approx. 0.3~mbar) has a linear effect on the detected count rate. This approximation is an upper limit for the influence of the residual gas on the surface measurement, since a larger gas volume contributes to the signal intensity in the transmission measurement. For both the C\nobreakdash-K edge and the O\nobreakdash-K edge, a factor of 0.5 is obtained. This means that at maximum, half of the signal height of the two features B in the C\nobreakdash-K edge spectrum and the O\nobreakdash-K edge spectrum can be attributed to the excited EtOH from the residual gas. Conversely, the adsorption of EtOH molecules on the surface is confirmed though our measurements. 
To further optimize the ratio between adsorption and gas signal for future measurements with the gas cell, the following steps can be adopted: on the one hand the cell volume can be flushed with an inert gas to minimize the residual gas in the cell. This step can be monitored by the measurement of the gas transmission. On the other hand the evacuation of the gas cell has to be further improved, e.g. by enlarging the pipe cross-sections. Furthermore, it should be noted for a suitable pressure determination that the sensitivity factors of the pressure measuring tubes used are not optimized for different gas compositions requiring dedicated calibrations and, thus, the pressure data is best used as a relative measurement. 
\section{\label{sec:summary}Summary and Outlook}
A measurement cell for the adsorption behavior of gases was designed and brought into operation. The design of the cell was adjusted to allow for investigation of the adsorption behavior of volatile organic compounds on various substrates. The cell is meant for usage in the soft x-ray range especially for the analysis of C, N and O and is constructed for flow-through operation in a high-vacuum chamber. The gas-cell allows the excitation of a surface under total reflection geometry by detecting the characteristic x-ray fluorescence lines. It can also be used for transmission experiments with simultaneous fluorescence detection.  This results in a flexible tool for the investigation of low-Z elements in a gaseous phase and their adsorption behavior to surfaces by using x-ray absorption spectroscopy in fluorescence and in transmission detection mode. 
For future experiments this measurement cell can be used with an updated evacuation for less residual gas in the surface mode for an absolute quantification of the absorbed molecules and in the transmission mode of operation it is conceivable to determine atomic fundamental parameter of light atoms in the gas phase.
\begin{acknowledgments}
This work was funded through the European Metrology Research Programme (EMRP) Project ENV56 KEY-VOCs. \cite{noauthor_emrp_2022} The EMRP is jointly funded by the EMRP participating countries within EURAMET and the European Union.\\
\end{acknowledgments}

\nocite{*}

\bibliography{Gaszelle_Final}

\end{document}